\documentclass[aps,prb,showpacs,onecolumn,superscriptaddress,preprint]{revtex4-1}
\usepackage{graphicx}
\usepackage{array}
\usepackage{amsmath,amsfonts,amssymb}
\usepackage[ansinew]{inputenc}
\usepackage{color}
\usepackage{hyperref}

\begin{document}

\title{Strong electron-phonon coupling in the $\sigma$ band of graphene}
\author{Federico Mazzola}
\affiliation{Department of Physics, Norwegian University of Science and Technology (NTNU), N-7491 Trondheim, Norway}

\author{Thomas Frederiksen}
\affiliation{\mbox{Donostia International Physics Center (DIPC) -- UPV/EHU,
E-20018 San Sebasti\'an, Spain}}
\affiliation{IKERBASQUE, Basque Foundation for Science, E-48013, Bilbao, Spain}

\author{Thiagarajan Balasubramanian}
\affiliation{MAX IV Laboratory, PO Box 118, S-22100 Lund, Sweden}

\author{Philip Hofmann}
\affiliation{Department of Physics and Astronomy, Interdisciplinary Nanoscience Center (iNANO), Aarhus University, Denmark}

\author{Bo Hellsing}
\affiliation{\mbox{Donostia International Physics Center (DIPC) -- UPV/EHU,
E-20018 San Sebasti\'an, Spain}}
\affiliation{Material and Surface Theory Group, Department of Physics,
University of Gothenburg, Sweden}

\author{Justin W. Wells}
\email[]{justin.wells@ntnu.no}
\affiliation{Department of Physics, Norwegian University of Science and Technology (NTNU), N-7491 Trondheim, Norway}
\begin{abstract}

First-principles studies of the electron-phonon coupling in graphene predict a high coupling strength for the $\sigma$ band with $\lambda$ values of up to 0.9. Near the top of the $\sigma$ band, $\lambda$ is found to be $\approx 0.7$. This value is consistent with the recently observed kinks in the $\sigma$ band dispersion by angle-resolved photoemission. While the photoemission intensity from the $\sigma$ band is strongly influenced by matrix elements due to sub-lattice interference, these effects differ significantly for data taken in the first and neighboring Brillouin zones. This can be exploited to disentangle the influence of matrix elements and electron-phonon coupling. A rigorous analysis of the experimentally determined complex self-energy using Kramers-Kronig  transformations further supports the assignment of the observed kinks to strong electron-phonon coupling and yields a coupling constant of $0.6(1)$, in excellent agreement with the calculations.

\end{abstract}

\date{\today}

\maketitle

The electron-phonon coupling (EPC) in graphene has been the subject of numerous studies \cite{Calandra:2007aa,Park:2008c,Ulstrup:2012,Bianchi:2010,Fedorov:2014}. Most of the literature focuses on the EPC in the $\pi$ band, as these states form the Fermi surface and the EPC thus directly affects the materials' transport properties \cite{Grimvall:1981,Hellsing:2002,Hofmann:2009ab}. EPC in the $\sigma$ band can be expected to be stronger than in the $\pi$ band for several reasons: The atomic orbital overlap for the $\sigma$ bands is substantially stronger than for the $\pi$ band and the $\sigma$ bands will thus be more sensitive to a vibration-related change of the bond length. Also, the $\pi$ band's EPC is quite special because of the vanishing density of states near the Dirac point and the accompanying phase space reduction. While the EPC in the $\sigma$ band has no direct implication for the transport properties of graphene, similar physics plays an important role in the superconductivity of the related material MgB$_2$ \cite{Choi:2002ab}. 

While no theoretical investigations have so far been published on the EPC in the $\sigma$ band, two recent angle-resolved photoemission (ARPES) studies come to entirely different conclusions based on very similar data. Mazzola \textit{et al.}\cite{Mazzola:2013a} have reported the observation of a kink-feature near the top of the $\sigma$ band and ascribed this to strong EPC with a mass enhancement parameter $\lambda$ between 0.7 and 1, depending on the graphene system. Similar kinks are often observed near the Fermi level and not usually expected and at higher binding energy. To explain the presence of the kink, Mazzola \textit{et al.} needed to assume that the EPC in the $\sigma$ band is determined by scattering effects involving predominately other $\sigma$ states. More recently, Jung \textit{et al.}\cite{Jung:2016} have reported similar data but have interpreted the observed kink in terms of strong matrix element effects which suppress the photoemission intensity near normal emission, without the need to envoke any EPC, i.e.,~essentially using $\lambda=0$. The difference in these interpretations does not only leave the question of the EPC strength open, it is also interesting in connection with the observation of other controversial kink-like features at higher binding energy \cite{Graf:2007, Valla:2007, Zhang:2008}.

In this Letter we present a calculation of the  EPC in the $\sigma$ band from first principles, yielding an energy-dependent coupling strength. The calculation gives detailed insight into the origin of the EPC and predicts its strength. We also report new ARPES results that have not been taken in the first Brillouin zone (1st BZ) near normal emission, as in the previous works, but in a neighboring Brillouin zone (NBZ), such that the matrix elements effects no longer suppress the emission from the top of the $\sigma$ band, qualitatively illustrating that the kink is not caused by matrix element effects. We also determine the electronic self-energy from the ARPES data and show, using Kramers-Kronig (KK) analysis, that the result is self-consistent and agrees with the theoretical prediction for the EPC. 

\begin{figure*}
\includegraphics[width=\textwidth]{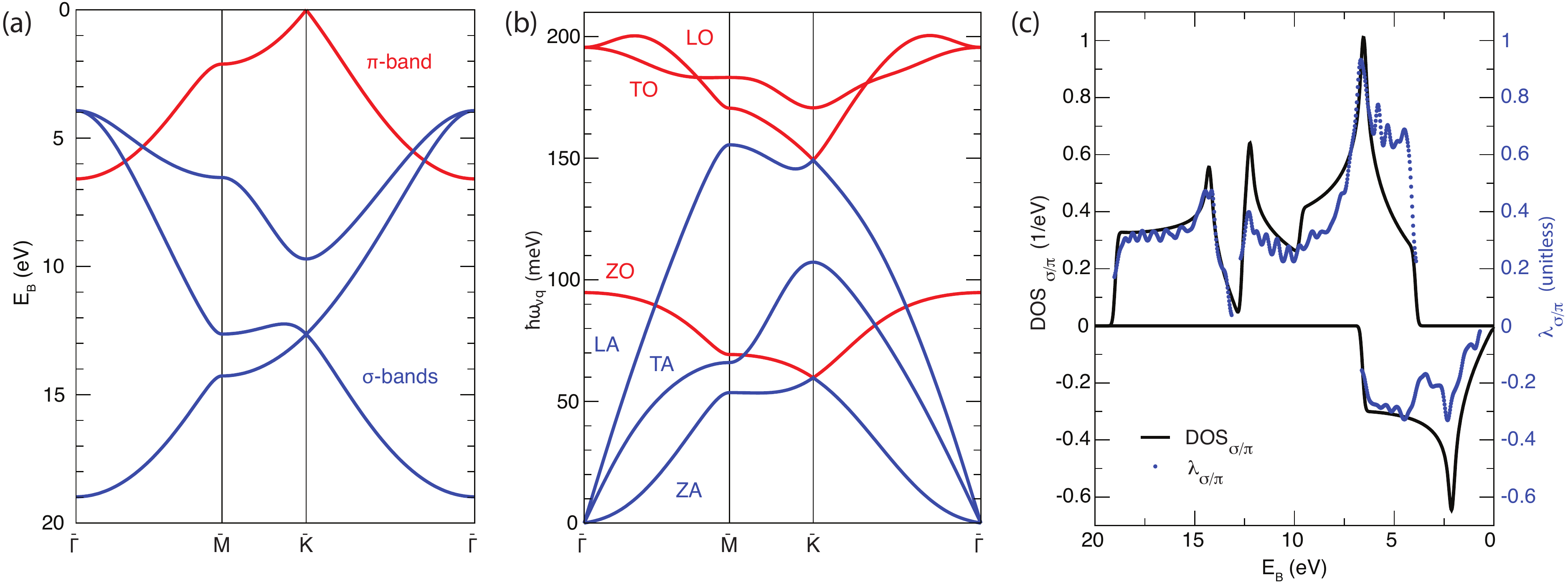}
\caption{(Color online) (a) Electronic band structure of graphene with the $\pi$ ($\sigma$) bands in red (blue). (b) Phonon band structure of graphene. The optical (acoustic) bands are shown by red (blue) lines.  (c) Electronic DOS (black lines) and EPC strength $\lambda_{\sigma/\pi}$ (blue points) for a photohole generated in either a $\sigma$- or $\pi$-band. The $\pi$-components are shown as negative for clarity. Near the top of the $\sigma$ band $\lambda_{\sigma}\approx0.7$.} 
\label{fig:1}
\end{figure*}

Calculations were based on Kohn-Sham density functional theory using the implementation in \textsc{Siesta} \cite{SoArGa.02.SIESTAmethodab} together with \textsc{Inelastica}\cite{FrPaBr.07.Inelastictransporttheory} for the EPC
\footnote{We performed supercell calculations with a $9\times9$ repetition of the primitive 2-atom graphene cell ($N=162$ atoms in total), using the PBE-GGA functional \cite{PeBuEr.96.Generalizedgradientapproximation}, a 400 Ry cutoff for the real-space grid, a SZ basis set with an 0.02 Ry energy shift for the cutoff radii, and a $4\times4$ $k$-point sampling in the self-consistency loop. The interatomic distance was 1.48 {\AA}. Force constants and gradients of the Kohn-Sham Hamiltonian were computed from finite differences with an amplitude of 0.02 {\AA}. $\rho$ and $\lambda$ were evaluated with a Gaussian smearing of 0.1 eV for the $\delta$-functions, a dense $18\times18$ $\mathbf{k}$-grid and $\Gamma$ phonons (i.e., $3N-3$ modes) for the shrunk Brillouin zone of the $9\times9$ supercell.} 
yielding the electronic band structure and the phonon dispersion relations [Fig.~\ref{fig:1}(a),(b)] of graphene in excellent agreement with previous results  \cite{WiRu.04.phonondispersiongraphite,MaReTh.04.PhononDispersionin,YaRuCh.08.Phonondispersionsand,WeGrLi.08.Phonon-MediatedTunnelinginto,LiZhBr.10.PhononInducedGaps}. In the low temperature limit, which is relevant for the ARPES experiments \cite{Mazzola:2013a,Jung:2016}, the thermal energy ($\approx 8$~meV) is less than the typical optical phonon energy ($\approx 170$~meV). In this case EPC by phonon emission  dominates while phonon absorption is suppressed. The Eliashberg function $\alpha^2F$ can then be written as
\begin{eqnarray}
\label{eq:elia_1} 
\alpha^2F^E_{n\mathbf k}(\omega) &=& \sum_{\nu \mathbf q} \sum_{n'\neq n}\; |g^{\nu}(n\mathbf k,n'\mathbf {k+q})|^2  
\delta(\varepsilon_{n'\mathbf {k+q}}-\varepsilon_{n\mathbf k}-\hbar\omega_{\nu\mathbf q})
  \delta(\hbar\omega-\hbar\omega_{\nu\mathbf q}),  
\end{eqnarray}
where the summation includes all electron scattering events 
from states $\varepsilon_{n'\mathbf {k+q}}$ into the photo-hole state $\varepsilon_{n\mathbf k}$ with emission of a phonon with an energy $\hbar\omega_{\nu\mathbf q}$, and mediated by the matrix elements $g^{\nu}(n\mathbf k,n'\mathbf {k+q})$. 
The  EPC parameter for the electronic state $n\mathbf k$ is defined as 
\begin{eqnarray}
\lambda_{n\mathbf k} &=& 2 \int d\omega \frac{\alpha^2 F^E_{n\mathbf k}(\omega)}{\omega} \nonumber \\
&=& \sum_{\nu \mathbf q} \sum_{nn'} \frac{2}{\hbar\omega_{\nu\mathbf q}}\; |g^{\nu}(n\mathbf k,n'\mathbf {k+q})|^2  \label{eq:lambda-nk}
\delta(\varepsilon_{n'\mathbf {k+q}}-\varepsilon_{n\mathbf k}-\hbar\omega_{\nu\mathbf q}). 
\label{eq:lambda-n-k}
\end{eqnarray}
Note that here $\lambda_{n\mathbf k}$ is a quantity depending on the energy of the electronic state, it does not  correspond to the mass enhancement parameter at the Fermi energy \cite{Gayone:2003aa}.  In view of the nearly isotropic band structure, we average $\lambda_{n \mathbf k}$ along the 2D constant energy contour \(\varepsilon_{n \mathbf k}=\varepsilon \), and sum up the contributions from the intraband and interband scattering within the $\sigma$- and $\pi$-bands, i.e., we \emph{define} the quantities
\begin{eqnarray}
\lambda_{\sigma}(\varepsilon) &\equiv& \frac{2}{\rho_{\sigma}(\varepsilon)}
\sum_{\sigma \mathbf k} \lambda_{\sigma \mathbf k}\delta(\varepsilon- \varepsilon_{\sigma \mathbf k}), \\
\lambda_{\pi}(\varepsilon)  &\equiv& \frac{2}{\rho_{\pi}(\varepsilon)}
\sum_{\pi \mathbf k} \lambda_{\pi \mathbf k}\delta(\varepsilon- \varepsilon_{\pi \mathbf k}),
\end{eqnarray}
where $\rho_{\sigma}(\varepsilon)$ and $\rho_{\pi}(\varepsilon)$ are the density of states (DOS) of the $\sigma$ and $\pi$ bands, respectively. Figure \ref{fig:1}(c) shows the calculation of these EPC parameters in the energy range corresponding to all occupied states. Near the Fermi energy the EPC is very small ($\lambda_\pi< 0.1$) and consistent with values of the order 0.1-0.3 reported for $n$- and $p$-doped graphene.\cite{Calandra:2007aa,JoUlBi.13.Electronphononcoupling} 
On the other hand, near the top of the $\sigma$ band we find a large value $\lambda_\sigma\approx 0.7$. This confirms the expectation that the EPC is considerably stronger in the $\sigma$ band than in the $\pi$-band.

\begin{figure*}
\includegraphics[width=0.9\textwidth]{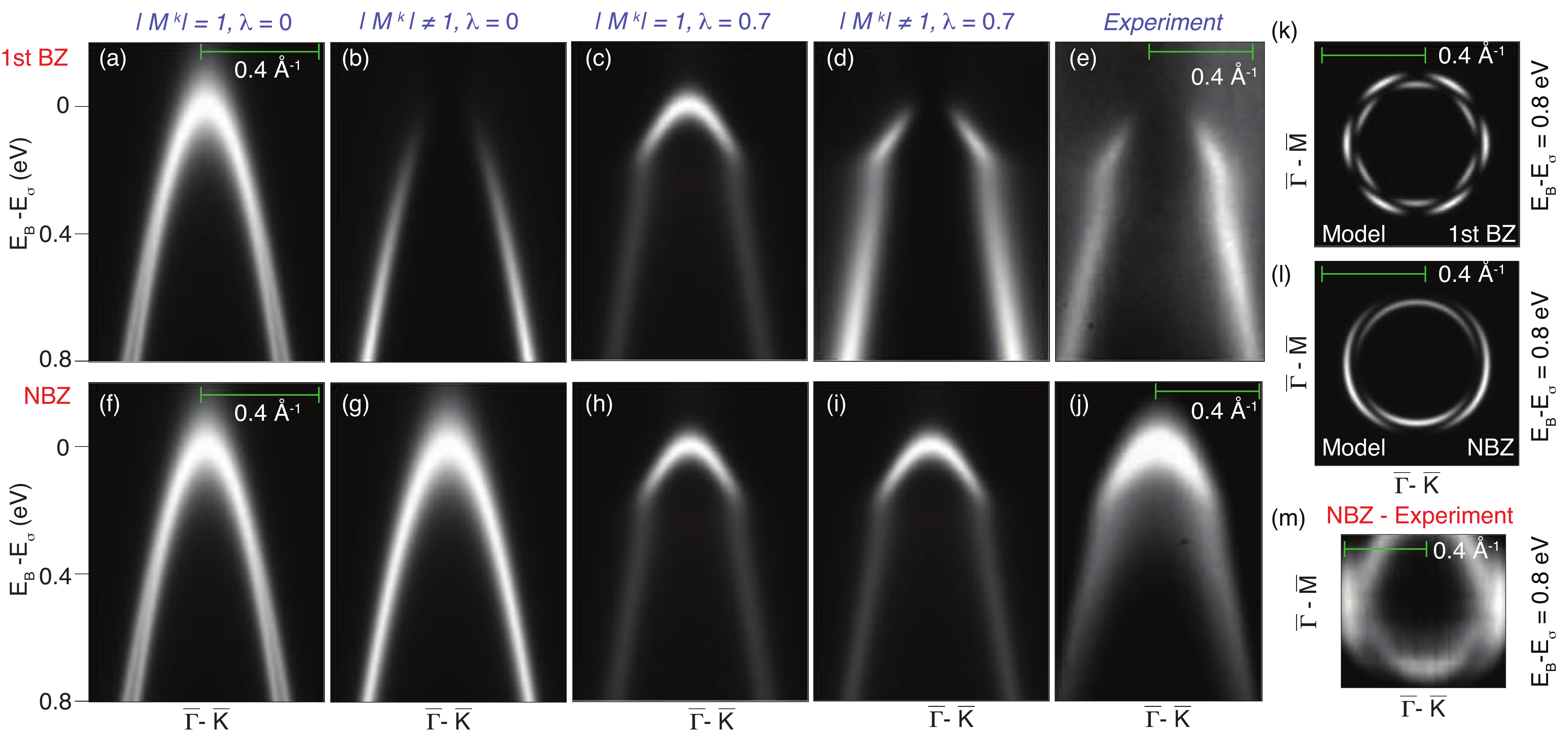}
\caption{(Color online) Effect of sub-lattice interference and EPC near the top of the $\sigma$-band. (a)-(e) Spectra relative to the first Brillouin zone (1st BZ) center. (f)-(j) Spectra relative to the center of the first neighboring Brillouin zone (NBZ). (a),(f) Spectral function determined using a tight-binding approach and a constant
$\Im\Sigma$ (with $\Re\Sigma=0$). (b),(g) Expected ARPES intensity without EPC, i.e. spectral function times calculated matrix elements to account for sub-lattice interference. (c),(h) Expected ARPES intensity with EPC but without sub-lattice interference. (d),(i) Expected ARPES intensity including both EPC and sub-lattice interference; (e),(j) Measured ARPES
intensities. (k),(l) Calculated ARPES intensity at a constant energy below the band maximum in the 1st and neighboring BZ respectively, not including EPC. (m) Experimental ARPES intensity in the neighboring BZ. }
\label{fig:angles}
\end{figure*}

Previously reported ARPES results for the $\sigma$ band \cite{Mazzola:2013a,Jung:2016} were taken near normal emission, in the 1st BZ of graphene, and revealed a pronounced kink near the band maximum. In this geometry, the interference involving the two atoms in the unit cell of graphene leads to a strong suppression of the photoemission intensity near the top of the $\sigma$ band \cite{Shirley:1995aa} and for bands purely comprised of $s$-states \cite{Lizzit:2010aa}. While the observation of a kink near the band maximum was not disputed, its origin was: Mazzola \textit{et al.}\cite{Mazzola:2013a} ascribed the kink to strong EPC. They were able to construct a model spectral function in good agreement with the experimental data, assuming that the sudden increase in the electron density of states could play a role equivalent to the Fermi-Dirac function cutoff at the Fermi level, hence reproducing the same physics at higher binding energies \cite{Mazzola:2013a}. Jung \textit{et al.}\cite{Jung:2016} showed that the intensity of the two $\sigma$-type sub-bands near $\bar{\Gamma}$ is strongly anisotropic, something that they argued could potentially lead to a kink induced by ``switching'' the photoemission intensity from one sub-band to another. A spectral function based on the sub-lattice interference, however, could not fit the data without the additional assumption of a strong change of the state's photoemission cross section over a small $k$ or energy range.  

The ARPES experiments reported here (performed on the same graphene-on-SiC sample described previously\cite{Mazzola:2013a,Mazzola:2015}) avoid the complication of the vanishing intensity near the $\sigma$ band top by taking data in the NBZ where no such total suppression occurs (for a calculation of the matrix elements see the supplementary information \cite{suppl}). ARPES experiments were carried out at the beamline I4 MAX-lab, Lund, Sweden\cite{Jensen:1997}, using linear-horizontal light polarisation. The sample temperature was $100$~K. The energy and momentum resolutions were better than $35$~meV and $0.018$~\AA$^{-1}$, respectively.

Figure \ref{fig:angles} illustrates the effect of collecting data in different geometries on the observation of the kink. Figure \ref{fig:angles}(a)-(e) and (f)-(j) show the situation in the 1st BZ and the NBZ, respectively, while Fig.~\ref{fig:angles}(a),(f) display a model spectral function for the $\sigma$ band maximum based on a simple first nearest neighbors tight-binding calculation \cite{suppl}. The images are identical, as this initial state dispersion is obviously periodic in reciprocal space. The striking role of the matrix elements becomes evident in Fig. \ref{fig:angles} (b),(g) which show the expected photoemission intensities, calculated using equations (1) and (2) from Ref. \onlinecite{Jung:2016} for the matrix elements $M^\mathbf{k}$ and the photoemision intensity, respectively, with energy-independent photoionization cross sections $A_s=0.5A_p$ for the first BZ ($h\nu=36$~eV), and $A_s=1.5A_p$ for the NBZ ($h\nu=75$~eV) \cite{Yeh:1985}. In the 1st BZ the photoemission intensity is totally suppressed near $\bar{\Gamma}$ but in the NBZ it is not. Note that this simulation does not show any kinks, despite of the inclusion of sub-lattice interference via the matrix elements. The effect of strong EPC is probed in Fig. \ref{fig:angles} (c),(h) and (d),(i). In (c),(h), the expected photoemission intensity is shown for $\lambda=0.7$ (calculated using a similar procedure as in Ref. \onlinecite{Mazzola:2013a} and further described in Ref.\ \onlinecite{suppl}) but the interference effects are switched off by setting the matrix elements $M^\mathbf{k}=1$. The strong kink is evident. Figure \ref{fig:angles}(d),(i) show the same calculation without artificially holding $M^\mathbf{k}=1$, thus the interference effect is recovered. In the 1st BZ the intensity is missing in the centre of the image but the kink is still evident. In the NBZ, the full dispersion including kink is visible. Figure \ref{fig:angles}(e),(j) show the corresponding experimental data which is in excellent agreement with Fig. \ref{fig:angles} (d),(i). This shows that the kink cannot be explained by sub-lattice interference without EPC. 

While the matrix elements do not suppress the photoemission intensity near the top of the $\sigma$ band in the NBZ, the intensity of the two sub-bands still remains unequal. Calculations of this are shown in Fig.~\ref{fig:angles}(k) and (l) for the 1st BZ and NBZ, respectively. The 1st BZ results agree with Ref.\ \onlinecite{Jung:2016}. The NBZ results show a two-fold symmetry with a much larger overlap between the sub-bands for certain angles. The results agree well with the experimental angular distribution in the NBZ shown in Fig.~\ref{fig:angles}(m) \cite{suppl}. 

The highly anisotropic matrix elements in the 1st BZ and the possibility to suppress one of the sub-bands completely can be exploited for a more quantitative analysis of the EPC because it removes the difficulty of fitting two bands. We use this for an alternative proof that the kink is caused by EPC by extracting the bare band dispersion along with  the real and imaginary parts of the electronic self-energy, $\Re \Sigma $ and $\Im \Sigma$.

\begin{figure}
\includegraphics[width=0.6\columnwidth]{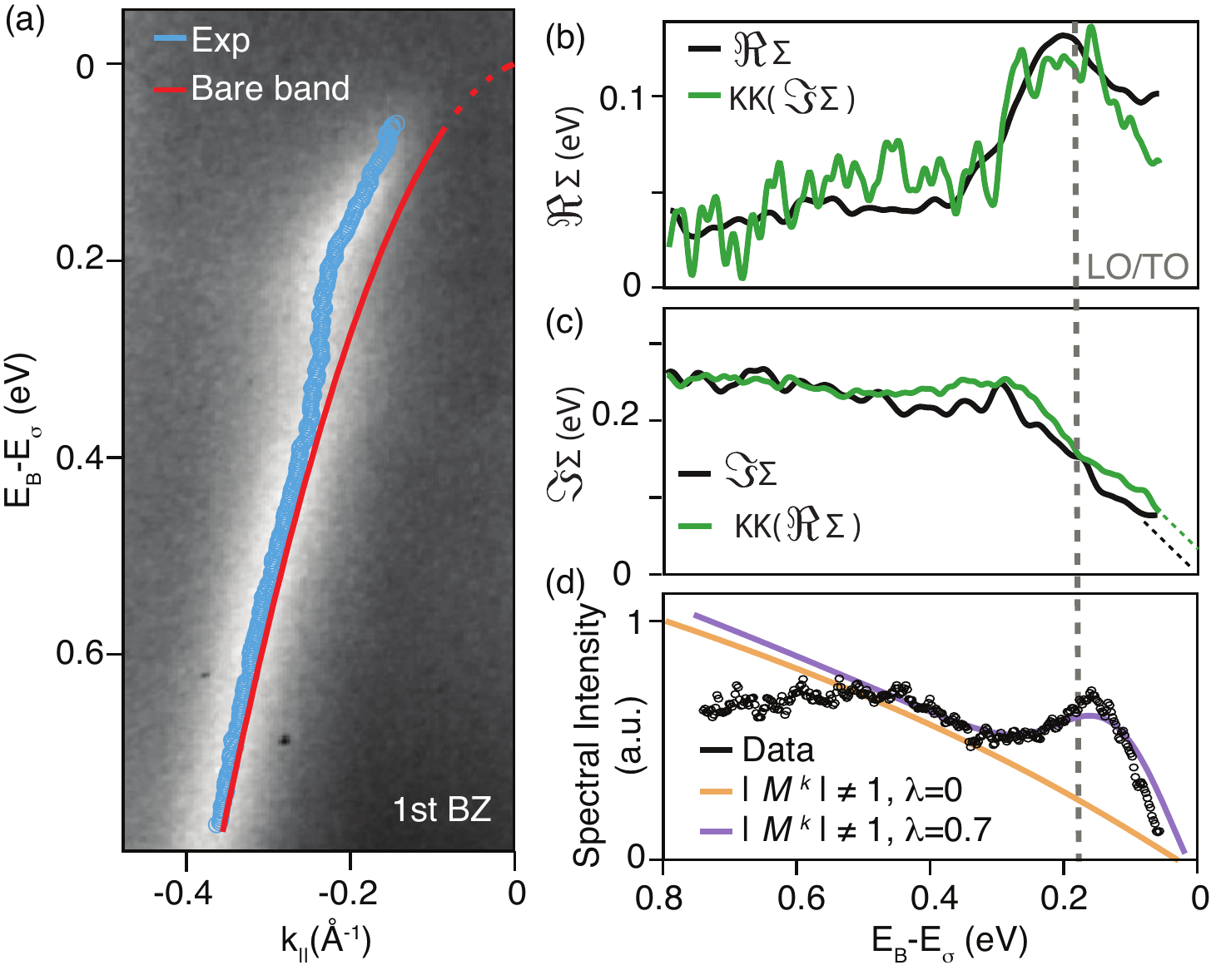}
\caption{(Color online) Analysis of the real and imaginary parts of the quasiparticle self-energy, $\Re\Sigma$ and $\Im\Sigma$. (a) ARPES data with the bare band and experimentally determined dispersion. (b) $\Re\Sigma$ (black) plotted alongside the KK-transformed $\Im\Sigma$ (green); (c) $\Im\Sigma$ (black), alongside the KK-transformed $\Re\Sigma$ (green). (d) Comparison between the calculated and experimentally determined spectral intensity (i.e. MDC  peak height) as function binding energy: the black curve is extracted from the experiment (a), the yellow curve from a simulated spectrum with inclusion of matrix elements but zero EPC and the purple curve is extracted from a simulated spectrum with inclusion of matrix elements and EPC.}
\label{KK}
\end{figure}

In principle, $\Re\Sigma$ and $\Im\Sigma$ can be determined independently from the measured spectral function but only when the bare band dispersion is known\cite{GaKiWe.05.Determiningelectronphonon}. Using a tight-binding model for this is not an adequate approach, since the parameters are not known with sufficient accuracy. We instead use the self-consistent method proposed by  Pletikosi\'c \textit{et al.}\cite{Pletikosic:2012} to extract  the bare band dispersion, $\Re\Sigma$ and $\Im\Sigma$.  Figure \ref{KK}(a) gives the ARPES data in the 1st BZ with the bare band dispersion (red) and the experimentally determined dispersion from momentum distribution curve (MDC) analysis (blue).  $\Re\Sigma$ is extracted from the experimentally determined dispersion relative to the bare band (i.e.\ the renormalisation) and plotted in black in Fig.\ \ref{KK}(b). $\Im\Sigma$ is extracted from the MDC linewidth and plotted in black in Fig.\ \ref{KK}(c). The kink is particularly well seen in $\Re\Sigma$. In order to confirm that the kink is due to EPC, we KK transform both $\Re\Sigma$ and $\Im\Sigma$ [referred to as KK$(\Re\Sigma)$ and KK$(\Im\Sigma)$, respectively] and plot the results in green in Fig.\ \ref{KK}(c) and (b), respectively. The similarity of $\Re\Sigma$ with KK$(\Im\Sigma)$ and $\Im\Sigma$ with KK$(\Re\Sigma)$ is striking. In all cases, it is also clear that the binding energy of the kink is at $\approx 200$ meV below the $\sigma$-band maximum, consistent with coupling to the LO and TO phonons. This analysis yields an EPC strength of $\lambda\approx0.6$, which is extracted following the method described in Refs.\ \onlinecite{Pletikosic:2012,JoUlBi.13.Electronphononcoupling} and is consistent with the calculated values.

Finally, we emphasize that the results here can be viewed to be consistent with those of Jung \textit{et al.}\cite{Jung:2016} The experimental data is very similar and the authors, after introducing the sub-lattice interference effect, find that the spectral function cannot be fitted within this model without the \emph{ad hoc} assumption of a photoemission cross section $A_s$ that is strongly $k$-dependent (or, equivalently, energy-dependent), so as to give rise to a `singularity' in $M^\mathbf{k}$ at the location of the kink [as shown in Fig.\ 4(a) of Ref.\ \onlinecite{Jung:2016}]. In the presence of EPC, such assumptions are not necessary because the  EPC anyway acts to redistribute the spectral intensity: Near the top of the band, the increased lifetime of the photohole leads to narrower MDCs with higher peak intensity values. Figure \ref{KK}(d) shows the experimentally determined intensity peak height (each MDC is fitted with a Voigt function, from which the peak height is found) alongside the peak height calculated both with and without EPC [extracted from Fig.\ \ref{fig:angles}(d) and (b), respectively]. Ignoring EPC gives rise to a spectral intensity which is smoothly increasing from zero at the energy of the band maximum, whereas the inclusion of EPC gives rise to a spectral intensity which is peaked at an energy $\approx170$ meV from the band maximum (corresponding to the energy of the LO/TO phonons).

 In conclusion, we investigated the EPC in graphene by first principles calculations and, for the $\sigma$ band, by ARPES investigations. The calculations predict high values of $\lambda$ near the $\sigma$ band maximum in excellent agreement with the experimental results. We show that the sub-lattice effect on the photoemission matrix effects has little relevance for the kink observed in ARPES, even though it influences the total intensity and the visibility of the two sub-bands. The interference-induced total suppression of a given sub-band can even be used for a quantitative analysis of the self-energy $\Sigma$ and the result shows a consistent picture of the dispersion kink being caused by EPC. \\

\textbf{Acknowledgments:} 
We thank Rositsa Yakimova for supplying the sample and Keun Su Kim, Kai Rossnagel, Mads Brandbyge, Tue Gunst, and Antti-Pekka Jauho for helpful discussions. We gratefully acknowledge funding from VILLUM FONDEN, Aarhus University Research Foundation, the Danish Council for Independent Research, Natural Sciences under the Sapere Aude program (Grant No.~DFF-4002-00029), Center for Nano\-structured Graphene (Project DNRF58), the Basque Departamento de Educaci\'on and the UPV/EHU (Grant No.~IT-756-13), the Spanish Ministerio de Econom\'ia y Competitividad (Grant No.~MAT2013-46593-C6-2-P), and the European Union FP7-ICT project PAMS (Contract No.~610446).


\bibliography{Mazzola.bib}

\end{document}